\documentclass[journal,12pt,draftclsnofoot,onecolumn]{IEEEtran}
\usepackage{color}
\usepackage{balance}
\usepackage{graphicx}
\usepackage{enumerate}
\usepackage{subfigure}
\usepackage{algorithm}
\usepackage{algorithmic}
\usepackage{booktabs}
\usepackage{cite}
\usepackage{amssymb}
\usepackage{amsmath}
\usepackage{float}
\usepackage{float}
\usepackage{bm}
\begin{document}

\title{Intelligent Reflecting Surface Assisted Anti-Jamming Communications: A Fast Reinforcement Learning Approach
\vspace{15pt}}

 \linespread{1.1}{

\author{

  Helin~Yang,~\IEEEmembership{Member,~IEEE},
          Zehui~Xiong,    Jun~Zhao,~\IEEEmembership{Member,~IEEE},
  Dusit~Niyato,~\IEEEmembership{Fellow,~IEEE},
    Qingqing~Wu,~\IEEEmembership{Member,~IEEE},
H. Vincent Poor,~\IEEEmembership{Fellow,~IEEE},
and  Massimo Tornatore,~\IEEEmembership{Senior Member,~IEEE}

        \vspace{-20pt}

\thanks{\small{H. Yang,  Z. Xiong, J. Zhao, and D. Niyato are with the School of Computer Science and Engineering, Nanyang Technological University, Singapore 639798 (e-mail: hyang013@e.ntu.edu.sg,  zxiong002@e.ntu.edu.sg, junzhao@ntu.edu.sg, dniyato@ntu.edu.sg).}}

\thanks{ \small{Q. Wu  is with the Department of Electrical and Computer Engineering, National University of Singapore, Singapore 119260 (email:elewuqq@nus.edu.sg).}}

\thanks{ \small{H. V. Poor is with the Department of Electrical Engineering, Princeton University, Princeton, NJ 08544  (email: poor@princeton.edu).}}

\thanks{ \small{M. Tornatore is with the Department of Electronics,  Information and Bioengineering, Politecnico di Milano, 20133 Milan, Italy (email: massimo.tornatore@polimi.it).}}

}
}

\maketitle

 \linespread{1.20}{

\begin{abstract}
 \normalsize{
Malicious jamming launched by smart jammers can attack legitimate transmissions, which has been  regarded as one of the critical security challenges in wireless communications. With this focus,  this paper considers the use of an intelligent reflecting surface (IRS) to enhance anti-jamming communication performance and mitigate jamming interference by adjusting the surface reflecting elements at the IRS. Aiming to enhance the communication performance against a smart jammer, an optimization problem for jointly optimizing power allocation at the base station (BS) and reflecting beamforming at the IRS is formulated while considering quality of service (QoS) requirements of legitimate users. As the jamming model and jamming behavior are dynamic and unknown, a fuzzy win or learn fast-policy hill-climbing (WoLF¨CPHC) learning approach is proposed to jointly optimize the anti-jamming power allocation and reflecting beamforming strategy, where WoLF¨CPHC is capable of quickly achieving the optimal policy without the knowledge of the jamming model, and fuzzy state aggregation can represent the uncertain environment states as aggregate states. Simulation results demonstrate that the proposed anti-jamming learning-based approach can efficiently improve both  the IRS-assisted system rate and transmission protection level compared with existing solutions.  \vspace{5pt}
}

\end{abstract}

}
\vspace{-7pt}

\IEEEpeerreviewmaketitle
\section{Introduction}
\IEEEPARstart{D}{ue} to the inherent broadcast and open nature of wireless channels [1], [2], wireless transmissions are vulnerable to jamming attacks. In particular, malicious jammers can intentionally send jamming signals over the legitimate channels to degrade communication performance [1]-[3], which has been considered as  one   serious threat in wireless communications. Jamming-related investigations not only provide solutions into wireless security guarantees against jamming, but also offer insights on the vulnerabilities of existing systems. In this regard, lots of technologies have been recently presented to defend wireless security against jamming attacks, including  frequency hopping, power control, relay assistance, beamforming, and so on.

Frequency-hopping (FH) is a powerful and widely-adopted techniques where a wireless user is allowed to quickly switch its current operating frequency to other frequency spectrum, thereby avoiding potential  jamming attacks [4]-[6].  In [4], a mode-FH approach was presented to jointly utilize mode hopping  and conventional FH to  decrease bit error rate in the presence of jammers. A stochastic game  was developed to study the interaction process between jammer and legitimate user [5], where the jammer and the transmitter can act  as two contending players to achieve the optimal attack and defense policies, respectively. In [6], Hanawal $et~al$. proposed a joint FH and rate-adaptation scheme to avoid jamming attacks in the presence of a jammer, where the transmitter has the ability to escape the jammer by switching its operating channel, adjusting its rate, or both. However, FH may become ineffective  if smart jammers attack multiple channels simultaneously and it also requires extra spectrum resources to evade jammers.  Besides FH, power control is another commonly used technique [3], [7]-[12]. As an example, references [7] and [8] investigated a jammed wireless system where the system operator tries to control the transmit power to maximize system rate while guaranteeing quality-of-service (QoS) requirements of legitimate receivers. The authors in [3], [9]-[11] developed the novel power control strategies for the anti-jamming problem. Specifically, they leveraged game theory to optimize the power control policy of the transmitter against  jammers, and simulations were provided to verify the effectiveness of the proposed anti-jamming power control approaches. In addition, a jamming-resistant receiver was designed to improve the robustness of communication system against jamming in [12], and the optimal power control scheme  was also developed to improve the achievable rate.

Recently, cooperative communication using trusted relays has been proposed as one promising anti-jamming technique for improving the physical layer security [13]-[17]. To reduce the adverse effects of  jamming signals, a joint relay selection and beamforming problem was formulated in relay-aided systems [13], which was solved by applying semi-determined relaxation (SDR) technique. In [14] and [15], the joint cooperative beamforming and jamming designs considering  worst-case robust schemes were proposed to maximize the achievable rate under the imperfect channel state information (CSI) of a jammer. In [16] and [17], the  advanced anti-jamming schemes were proposed to cancel strong  jamming signal, with the goal to maximize the signal-to-interference-plus-noise-ratio (SINR) of legitimate users.

To deal with uncertain and/or  unknown jamming attack models, such as jamming policies and jamming power levels, reinforcement learning (RL) algorithms have been applied in some existing studies  to optimize the jamming resistance policy in dynamic wireless communication systems [18]-[22]. In [18], a policy hill climbing (PHC)-based Q-Learning approach was developed to improve the vehicular communication performance against  jamming without knowing the jamming model. Additionally, a fast PHC-based power control algorithm was also presented to assist base station (BS) to select anti-jamming transmit power strategy over multiple antennas [19]. As traditional Q-learning algorithm has slow convergence to the optimal policy, in [20] and [21], the authors adopted deep reinforcement learning (DRL) algorithms that enable transmitters to quickly obtain an optimal communication policy to guarantee security performance against smart jamming without the need of knowledge of the jammer's characteristics

However, despite the effectiveness of the above aforementioned anti-jamming schemes [3]-[21], employing  a  number  of  active  relays  incurs an  excessive  hardware  cost  and    system  complexity. Moreover, anti-jamming beamforming and power control in communication systems are generally  energy-consuming as more transmit power as well as circuit power  need to be consumed to improve the communication performance. To tackle these shortcomings, a new paradigm, called intelligent reflecting surface (IRS) [22]-[26], has been recently proposed as a promising technique to enhance  spectrum efficiency  and  secrecy performance in the fifth-generation (5G) and beyond communication systems. In particular, IRS is a uniform planar array  comprising of a large number of low-cost passive reflecting elements, where each of the elements adaptively adjusts its reflection amplitude and/or phase to control the strength and direction of the reflected electromagnetic wave, thus  enhancing and/or weakening the the received signals   at different users [23]-[26]. As a result, IRS has been employed in wireless commutation systems for security performance optimization [27]-[32]. In [27]-[31], the authors investigated the physical layer security enhancement of IRS-assisted communications systems, where both the BS¡¯s beamforming and the IRS¡¯s phase shifts were jointly optimized to improve secrecy rate in the presence of an eavesdropper. Additionally, two algorithms, called  alternative optimization (AO) algorithm and  SDP relaxation algorithm were applied to address the joint optimization problem in [27]-[30], and simulation results demonstrated the effectiveness of the presented IRS-assisted system compared with traditional systems. Furthermore, Yang $et~al$. in [32] applied DRL to learn the secure beamforming policy in multi-user IRS-aided secure systems, in order to maximize the system secrecy rate in the presence of multiple  eavesdroppers. To the best of our knowledge, IRS has not been explored yet in the existing works [4]-[32] to enhance the anti-jamming strategy against smart jamming, where the smart jammer attempts to deteriorate the quality of intended transmissions by transmitting jamming signal over the legitimate channels.

In this paper, we propose an IRS-assisted anti-jamming solution for  securing  wireless communications. In particular, we aim to maximize the system rate of multiple legitimate users in the presence of a smart multi-antenna jammer, while guaranteeing the QoS requirements of users against smart jamming. As the communication system environment is complex and the jammer is smart, a fuzzy RL based anti-jamming approach is proposed to effectively jointly optimize the anti-jamming power allocation and reflecting beamforming in uncertain environments. The main contributions are summarized as follows.

\begin{itemize}
\item  We, for the first time, propose a novel anti-jamming model based on the IRS-assisted system that jointly optimizes the transmit power allocation at the BS and the reflecting beamforming at the IRS to improve secrecy rate against smart jamming.
\item  Aiming to enhance the anti-jamming performance, an optimization problem for jointly optimizing power allocation and reflecting beamforming is formulated given QoS requirements of legitimate users. Since it is difficult to address the non-convex optimization problem in dynamic environments, the problem is modelled as an RL process.
\item  As the jamming model and jamming behavior are dynamic and unknown, a fuzzy  win or learn fast-policy hill-climbing (WoLF¨CPHC) learning approach is proposed to achieve the optimal anti-jamming  policy, where WoLF¨CPHC is capable of quickly achieving the optimal policy without knowing the jamming model, and fuzzy state aggregation (FSA) has the ability to represent the uncertain environment states as aggregate states.
\item  Simulation results are provided to validate that the IRS-assisted system significantly improves the anti-jamming  communication performance compared to conventional systems without IRS, and also to verify the effectiveness of the proposed learning approach in terms of improving the system rate and service protection level, compared with the existing approaches. For example, the proposed learning approach achieves the system rate and service protection level improvements of  21.29\% and 13.36\%, respectively, over the existing solution in [29].
\end{itemize}

The remainder of this paper is organized as follows. Section II provides the system model and problem formulation. Section III models the optimization problem as an RL process, and proposes  the fuzzy WoLF¨CPHC-based learning approach. Simulation results are provided in Section IV, and the paper is concluded in Section V.

\textit{Notations:} In this paper, vectors and matrices are represented by Boldface lowercase and uppercase letters, respectively. ${\rm{Tr}}( \cdot )$, ${( \cdot )^ * }$ and ${( \cdot )^H}$ respectively  stand for the trace, the conjugate and the conjugate transpose operations. $| \cdot |$ and  $|| \cdot ||$ denote  the absolute value of a scalar and the Euclidean norm of a vector or matrix, respectively. ${\mathbb{C}^{M \times N}}$ is the space of $ M \times N$ complex-valued matrices. $\mathbb{E}[ \cdot ]$ denotes the statistical expectation operation.

\section{System Model and Problem Formulation}

\subsection{System Model}

\begin{figure}
\centering

\includegraphics[width=0.6\columnwidth]{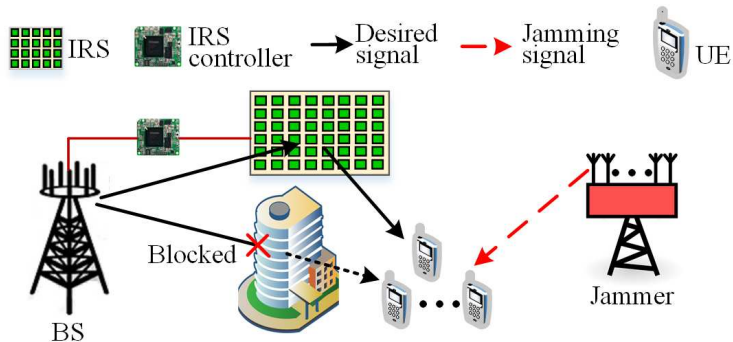}
\caption{{Illustration of an IRS-assisted communication system against a multi-antenna jammer.} } \label{fig:Schematic}

\end{figure}

As shown in Fig. 1, this paper considers an IRS-assisted communication system, which
consists of one BS with $N$ antennas and $K$ single-antenna
legitimate user equipments (UEs) located at the cell-edge.
The IRS comprised of $M$ reflecting elements is deployed to provide additional communication links so
as to improve the performance for the UEs over a given frequency
band. The direct communication links of cell-edge UEs may suffer high signal attenuation and
these links are severely blocked
by obstacles when these UEs are located in dead zones. In addition, as
illustrated in Fig. 1,  a malicious multi-antenna jammer is located near the
legitimate UEs who attempts to interfere the legitimate
transmissions by sending faked or replayed jamming signal for the
UEs via ${N_{\rm{J}}}$ antennas, in order to degrade the legitimate
communication performance. In this case, deploying the IRS can
effectively enhance the desired signal power and mitigate the
jamming interference generated from the jammer by designing the
reflecting beamforming at the IRS.

Let  $\mathcal{K} = \{ 1,2,...,K\} $ and $\mathcal{M} = \{
1,2,...,M\} $ represent the UE set and the IRS reflecting element
set, respectively.  Let ${\bf{G}} \in {{\mathbb{C}^{M \times N}}}$,
${\bf{g}}_{{\rm{bu,}}k}^H \in {\mathbb{C}^{1 \times N}}$,
${\bf{g}}_{{\rm{ru,}}k}^H \in {\mathbb{C}^{1 \times M}}$, and
${\bf{h}}_{{\rm{J,}}k}^H \in {\mathbb{C}^{1 \times {N_{\rm{J}}}}}$
denote the channel coefficients between the BS and the IRS, between the
BS and the $k$-th UE, between the IRS and the $k$-th UE, and between the
jammer and the $k$-th UE, respectively. The quasi-static flat-fading model is assumed for all the above channels. Let ${\bf{\Phi }} = {\rm{diag}}({\Phi
_1},{\Phi _2},....,{\Phi _M}) \in {\mathbb{C}^{M \times M}}$ denote the
reflection coefficient matrix associated with effective phase
shifts at the IRS, where ${\Phi _m} = {\omega _m}{e^{j{\theta
_m}}}$ comprises both  a reflection amplitude   ${\omega _m} \in [0,1]$ and
a phase shift coefficient  ${\theta _m} \in [0,2\pi ]$ on the
combined received signal. Since each phase shift is
favorable to be designed to achieve  maximum signal reflection,
we consider that ${\omega _m} = 1$, $\forall m \in \mathcal{M}$ in this paper [23]-[26].

The transmitted signal at the BS can be expressed as
\begin{equation}
\begin{split}
{\bf{x}} = \sum\nolimits_{k = 1}^K {\sqrt {{P_k}}
{{\bf{w}}_k}{s_k}}
\end{split}
\end{equation}
where ${P_k}$ stands for the transmit power allocated for the $k$-th
UE, and we have the power constraint: $\sum\nolimits_{k =
1}^K {{P_k}}  \le {P_{\max }}$ with ${P_{\max }}$ being the
maximum transmit power of the BS, ${s_k}$ is the transmitted symbol for the
$k$-th UE, ${s_k} \in \mathbb{C}$, $\mathbb{E}\{ {s_k}\}  = 0$ and
$\mathbb{E}{\rm{\{ |}}{s_k}{{\rm{|}}^{\rm{2}}}{\rm{\} }} =
{\rm{1}}$ which denotes the unit power information symbol, and
${{\bf{w}}_k} \in {\mathbb{C}^{N \times 1}}$ is the beamforming vector for
the $k$-th UE with ${\left\| {{{\bf{w}}_k}} \right\|^2} = 1$,
respectively.

This study considers the case that the smart jammer attempts to
disturb the BS's transmitted signal by emitting jamming signal
${{\bf{z}}_k} \in {\mathbb{C}^{{N_{\rm{J}}} \times 1}}$ at the $k$-th UE.  In addition, the
transmit power of the faked jamming signal for the $k$-th UE is
denoted as ${P_{{\rm{J}},k}} = {\left\| {{{\bf{z}}_k}} \right\|^2} =
{\rm{Tr}}({{\bf{z}}_k}{\bf{z}}_k^H)$. In this case, for
UE $k$, the received signal consists of the signal coming from its
associated BS, the reflected signal from the IRS and the jamming
signal from the jammer, which is written by
\begin{equation}
\begin{split}
{y_k} = \underbrace {\left( {{\bf{g}}_{{\rm{ru}},k}^H{\bf{\Phi }}{\bf{G}}
+ {\bf{g}}_{{\rm{bu}},k}^H} \right)\sqrt {{P_k}}
{{\bf{w}}_k}{s_k}}_{{\rm{desired}}\;{\rm{signal}}} + \underbrace
{\sum\limits_{i \in \mathcal{K},i \ne k} {\left(
{{\bf{g}}_{{\rm{ru}},k}^H{\bf{\Phi }}{\bf{G}} + {\bf{g}}_{{\rm{bu}},k}^H}
\right)\sqrt {{P_i}} {{\bf{w}}_i}{s_i}} }_{{\rm{inter-user
~interference}}} + \underbrace {\sqrt {{P_{{\rm{J}},k}}}
{\bf{h}}_{{\rm{J}},k}^H{{\bf{z}}_k}}_{{\rm{jamming}}\;{\rm{signal}}}
+ {n_k}
\end{split}
\end{equation}
where  ${n_k}$ denotes the additive complex Gaussian noise with the
zero mean and variance $\delta _k^2$  at the $k$-th UE. In (2), in
addition to the received desired signal, each UE also suffers
inter-user interference (IUI) and the jamming interference signal
in the system. According to (2), the received SINR at the $k$-th UE can be expressed as
\begin{equation}
\begin{split}
SIN{R_k} = \frac{{{P_k}{{\left| {\left(
{{\bf{g}}_{{\rm{ru}},k}^H{\bf{\Phi }}{\bf{G}} + {\bf{g}}_{{\rm{bu}},k}^H}
\right){{\bf{w}}_k}} \right|}^2}}}{{\sum\limits_{i \in \mathcal{K},i \ne k}
{{P_i}{{\left| {\left( {{\bf{g}}_{{\rm{ru}},k}^H{\bf{\Phi }}{\bf{G}} +
{\bf{g}}_{{\rm{bu}},k}^H} \right){{\bf{w}}_i}} \right|}^2}}  +
{P_{{\rm{J}},k}}{{\left| {{\bf{h}}_{{\rm{J}},k}^H{{\bf{z}}_k}}
\right|}^2} + \delta _k^2}}.
\end{split}
\end{equation}

\subsection{Problem Formulation}

In this paper, we aim to jointly optimize the transmit power
allocation  ${\{ {P_k}\} _{k \in \mathcal{K}}}$ at the BS and the reflecting
beamforming matrix ${\bf{\Phi }}$ at the IRS to maximize the
system achievable rate of all UEs against smart jamming,
subject to the transmit power constraint and the minimum  received SINR
constraints. Accordingly, the optimization problem can be formulated as
\begin{equation}
\begin{split}
\begin{array}{l}
\mathop {\max }\limits_{{{\{ {P_k}\} }_{k \in \mathcal{K}}},{\bf{\Phi }}} \; \sum\limits_{k \in \mathcal{K}} {{{\log }_2}} \left( {1 + SIN{R_k}} \right)\\
s.t.\;({\rm{a}}):\;\sum\nolimits_{k = 1}^K {{P_k}}  \le {P_{\max }},\\
\;\;\;\;\;\;({\rm{b}}):\;SIN{R_k} \ge SINR_k^{\min },\;\forall k \in \mathcal{K},\\
\;\;\;\;\;\;({\rm{c}}):\;|{\Phi _m}| = 1,\;0 \le {\theta _m} \le 2\pi
,\;\forall m \in \mathcal{M}
\end{array}
\end{split}
\end{equation}
where $SINR_k^{\min }$ denotes the minimum SINR  threshold
of the $k$-th UE. Note that  problem (4) is a non-convex optimization
problem, where the objective function is non-concave over the reflecting
beamforming matrix
${\bf{\Phi }}$; furthermore, the transmit power
allocation  variables  ${\{ {P_k}\} _{k \in
\mathcal{K}}}$ and ${\bf{\Phi }}$
  are intricately coupled in the objective function, thus rendering the joint optimization problem difficult to be solved optimally. So far, many optimization algorithms
(e.g., AO and SDR) have been proposed to obtain  an approximate
solution to problem  (4), by iteratively updating either
${\{ {P_k}\} _{k \in
\mathcal{K}}}$ or ${\bf{\Phi }}$   with the other fixed at each iteration. Hence, this paper proposes an
effective solution  to address such kind of the optimization problem, which will be provided in the next section. In addition, it is worth noting that this study mainly pays attention to jointly optimize the power allocation and the reflecting beamforming, so the transmit beamforming vector ${{\bf{w}}_k}$ is set by maximizing output SINR in (3) to make an easy presentation [17].

\section{Joint Power Allocation and Reflecting Beamforming Based on RL}

The problem formulated in (4) is difficult to be solved as
mentioned at the end of the last section. The traditional
optimization algorithms (e.g., AO and SDR) are capable of
addressing one single time slot optimization problem, but they may
achieve the suboptimal solution and achieve the greedy-search like
performance as the historical system information
and the long term benefit are ignored. In addition, the unknown jamming model
and channel variation result in dynamic open and uncertain characteristics,
which increases difficulty of solving the optimization problem.

Model-free RL is one of the dynamic programming tools which has the ability
to address the decision-making problem by achieving an optimal
policy in dynamic uncertain environments [33]. Thus, this paper
models the optimization problem as an RL, and a fuzzy
WoLF-PHC-based joint power allocation and reflecting beamforming
approach is proposed to learn the optimal anti-jamming strategy.

\subsection{Optimization Problem Transformation Based on RL}

In RL, the IRS-assisted communication system acts as an
environment and the central controller at the BS is regarded as a
 learning agent. In addition to the environment and the agent,
an RL also includes a set of possible system states $\mathcal{S}$,
a set of available actions $\mathcal{A}$, and a reward function
$r$, where the learning agent continually learns by interacting with the environment. The main elements
of RL are introduced as follows:

\textbf{States:} The system state ${s^t} \in \mathcal{S}$  is the
discretization of the observed information from the environment at
the current time slot $t$. The system state ${s^t}$ includes the
previous jamming power, i.e., ${\{ P_{{\rm{J}},k}^{t - 1}{\rm{\}
}}_{k \in \mathcal{K}}}$ according the channel quality, the
previous UEs' SINR values ${\{ SINR_k^{t - 1}{\rm{\} }}_{k \in
\mathcal{K}}}$, as well as the current estimated channel
coefficients ${\{{\bf{g}}_k^t{\rm{\} }}_{k \in \mathcal{K}}}$,
and it is defined as
\begin{equation}
\begin{split}
{s^t} = \left\{ {{{\{ P_{{\rm{J}},k}^{t - 1}{\rm{\} }}}_{k \in
\mathcal{K}}},{{\{{\bf{g}}_k^t{\rm{\} }}}_{k \in
\mathcal{K}}},{{\{ SINR_k^{t - 1}{\rm{\} }}}_{k \in \mathcal{K}}}}
\right\}.
\end{split}
\end{equation}

\textbf{Actions:}  The action  ${a^t} \in \mathcal{A}$ is one of
the valid selections that the learning agent chooses at the time
slot $t$, and it includes the transmit power  ${\{ {P_k}\} _{k \in
\mathcal{K}}}$ and the reflecting beamforming
coefficient (phase shift)  ${\{ {\theta _m}\} _{m \in
\mathcal{M}}}$. Hence, the action  ${a^t}$ is given by
\begin{equation}
\begin{split}
{a^t} = \left\{ {{{\{ P_k^t\} }_{k \in \mathcal{K}}},{{\{ \theta
_m^t\} }_{m \in \mathcal{M}}}} \right\}.
\end{split}
\end{equation}

\textbf{Transition probability:} $\mathcal{P}( \cdot )$ is a
transition model which represents the probability of taking an action
$a$ at a current state $s$ and then ending up in the next state
$s'$, i.e., $\mathcal{P}(s'|s,a)$.

\textbf{Policy:} Let $\pi ( \cdot )$ denote a policy that it maps
the current system state to a probability distribution over the
available actions which can be taken by the agent, i.e., $\pi (a,s):S
\to \mathcal{A}$.

\textbf{Reward function:} The reward function design plays an
important role in the policy learning in RL, where the reward
signal correlates with the desired goal of the system performance.
In the  optimization problem considered in Section
II.B, our objectives are threefold: maximizing the UEs' achievable
rate while decreasing the power consumption at the BS as much  as
possible and guaranteeing the SINR constraints against smart
jamming.

Based on the above analysis, the reward function is set as
\begin{equation}
\begin{split}
r = \underbrace {\sum\limits_{k \in \mathcal{K}} {{{\log }_2}}
\left( {1 + SIN{R_k}} \right)}_{{\rm{part}}\;{\rm{1}}} -
\underbrace {{\lambda _1}\sum\limits_{k \in \mathcal{K}} {{P_k}}
}_{{\rm{part}}\;2} - \underbrace {{\lambda _2}\sum\limits_{k \in
\mathcal{K}} {SINR_k^{{\rm{outage}}}} }_{{\rm{part}}\;3}
\end{split}
\end{equation}
where
\begin{equation}
\begin{split}
SINR_k^{{\rm{outage}}} = \left\{ \begin{array}{l}
0,\;{\rm{if}}\;SIN{R_k} \ge SINR_k^{\min },\forall k \in \mathcal{K},\\
1,\;{\rm{otherwise}}{\rm{.}}
\end{array} \right.
\end{split}
\end{equation}

In (7), the part 1 represents the immediate utility (system
achievable rate), the part 2 and part 3 are the cost functions
which are defined as the transmission cost of the power
consumption at the BS and the violation of minimum  SINR requirements,
respectively, with ${\lambda _1}$ and  ${\lambda _2}$ being the corresponding coefficient.  The goal of (8) is to impose the
SINR protection level, where the cost is zero (i.e.,
$SINR_k^{{\rm{outage}}} = 0$) when the SINR constraint is
guaranteed against jamming, otherwise,  $SINR_k^{{\rm{outage}}} =
1$.

The objective of the learning agent is to obtain an optimal policy
that optimizes the long-term cumulative discounted reward instead
of its immediate reward, which can be expressed as ${R_t} =
\sum\nolimits_{j = 0}^\infty  {{\gamma ^j}{r^{(t + j + 1)}}} $,
where $\gamma  \in (0,1]$ denotes the discount factor. Adopting
${Q^\pi }({s^t},{a^t})$ as the state-action value function, which
represents the value of executing an action $a$ in a state $s$
under a policy $\pi $,  it can be expressed as
\begin{equation}
\begin{split}
{Q^\pi }({s^t},{a^t}) = {E_\pi }\left[ {\sum\limits_{j = 0}^\infty
{{\gamma ^j}{r^{(t + j + 1)}}} |{s^t} = s,{a^t} = a} \right].
\end{split}
\end{equation}

Note that similar to [33], the state-action  Q-function
${Q^\pi }({s^t},{a^t})$ satisfies the Bellman equation which is  expressed
as
\begin{equation}
\begin{split}
{Q^\pi }({s^t},{a^t}) = {E_\pi }\left[ {{r^{t + 1}} + \gamma
\sum\limits_{{s^{t + 1}} \in \mathcal{S}} {P({s^{t +
1}}|{s^t},{a^t})\sum\limits_{{a^{t + 1}} \in \mathcal{A}} {\pi ({s^{t +
1}},{a^{t + 1}}){Q^\pi }({s^{t + 1}},{a^{t + 1}})} } } \right].
\end{split}
\end{equation}

The conventional Q-Learning algorithm is widely utilized to search
the optimal policy ${\pi ^ * }$. From (10), the optimal Q-function
(Bellman optimality equation) associated with the optimal policy
has the following form
\begin{equation}
\begin{split}
{Q^ * }({s^t},{a^t}) = {r^{t + 1}} + \gamma \sum\limits_{{s^{t +
1}} \in \mathcal{S}} {\mathcal{P}({s^{t + 1}}|{s^t},{a^t})} \mathop {\max
}\limits_{{a^{t + 1}} \in \mathcal{A}} {Q^*}({s^{t + 1}},{a^{t + 1}}).
\end{split}
\end{equation}

It is worth noting that the Bellman optimality equation generally
does not have any closed-form solution. Thus, the optimal
Q-function (11) can be solved recursively to achieve the optimal
${Q^ * }({s^t},{a^t})$ by using an iterative method. Accordingly,
the updating on the state-action value function $Q({s^t},{a^t})$
is expressed as
\begin{equation}
\begin{split}
\begin{array}{l}
Q({s^t},{a^t}) \leftarrow (1 - \alpha )Q({s^t},{a^t})\\
\;\;\;\;\;\;\;\;\;\;\;\;\; + \alpha \left( {{r^t} + \gamma \mathop
{\max }\limits_{{a^t} \in \mathcal{A}} {Q^*}({s^{t + 1}},{a^t})} \right)
\end{array}
\end{split}
\end{equation}
where   $\alpha  \in (0,1]$ stands for the learning rate for the update of
Q-function.

\subsection{Fuzzy WoLF-PHC-Based Joint Power Allocation and Reflecting
Beamforming}

Most of existing RL algorithms are value-based RL, such as
Q-Leaning, Deep Q-Network (DQN) and double DQN. These
RL algorithms can estimate the Q-function with low variance as well as
adequate exploration of action space, which can be ensured by
using the greedy scheme. In addition, policy gradient based RL
algorithm has the ability to tackle the continuous action space
optimization problems, but it may converge to suboptimal solutions
[33]. However, it is not easy to achieve the optimal anti-jamming
policy without knowing the jamming model and jamming strategy.

\begin{figure}
\centering
\includegraphics[width=0.6\columnwidth]{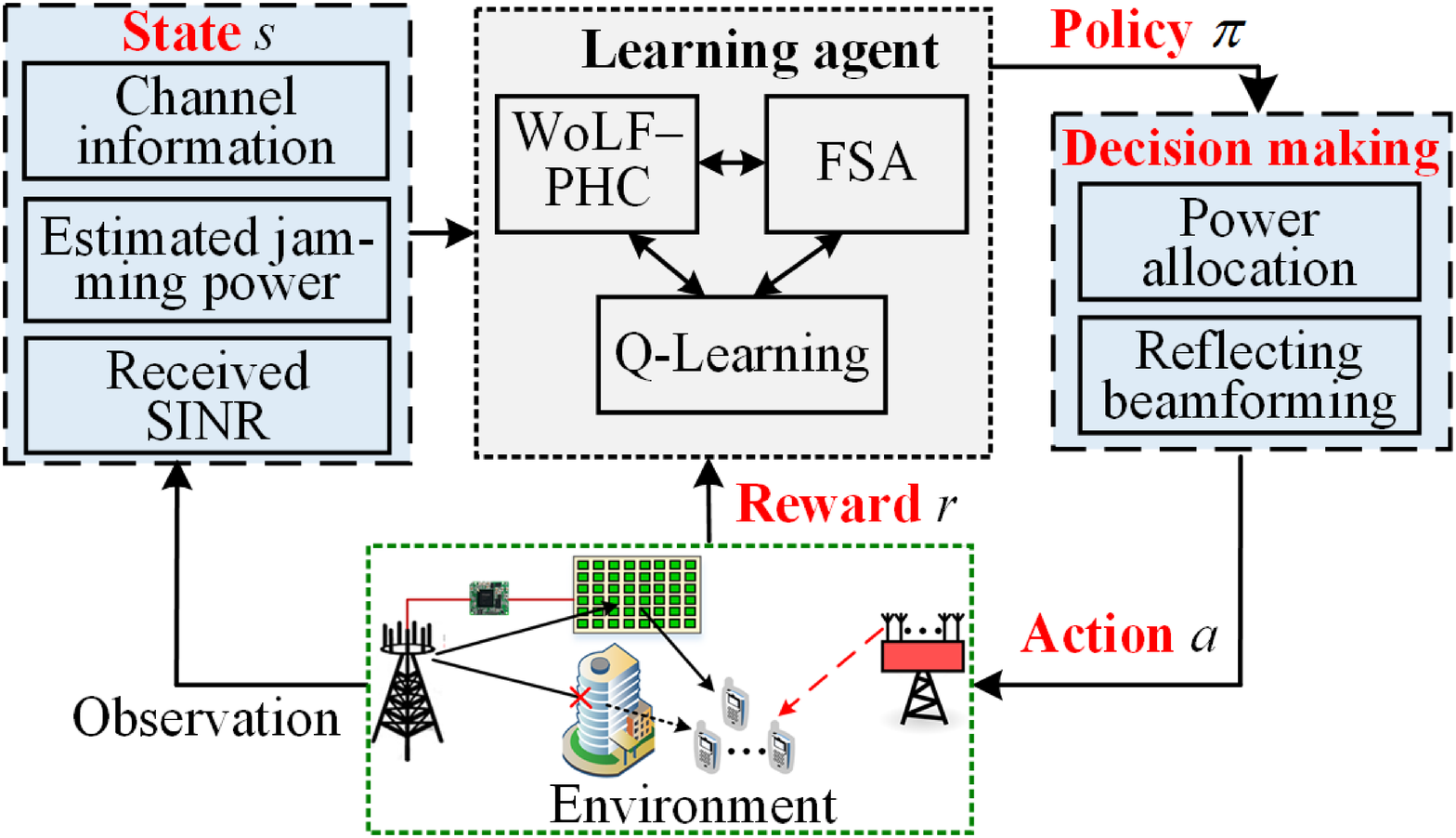}
\caption{{Fuzzy WoLF-PHC-based anti-jamming policy for
IRS-assisted systems.} } \label{fig:Schematic}
\end{figure}

In order to obtain the optimal anti-jamming policy against smart
jamming, we propose a fast fuzzy WoLF-PHC-based joint power
allocation and reflecting beamforming for IRS-assisted
communication systems, as shown in Fig. 2, where WoLF-PHC is
utilized to enable the learning agent to learn and adapt faster in
dynamic uncertain environments, and FSA is used to enable
Q-Learning to represent the system states with a fixed number of
aggregate states and represent continuous state spaces as
discrete. In the IRS-assisted system, the learning agent observes
a system state and receives an instantaneous reward by interacting
with the environment. Then, such information is leveraged  to train the
learning model to choose the anti-jamming policy with the maximum
Q-function value. After that, according to the selected policy,
the action is chosen to make decision in terms of power allocation
and reflecting beamforming. The procedures of the proposed
learning based decision making approach are provided in the
following analysis.

The $\varepsilon-$greedy policy is capable of balancing the tradeoff between an exploitation and an
exploration in an RL, in order to avoid
converging to local optimal power allocation and reflecting
beamforming strategy. In the $\varepsilon  - $greedy policy, the
agent selects the action with the maximum Q-table value with
probability  ${\rm{1}} - \varepsilon $, whereas a random action is
picked with probability  $\varepsilon $ to avoid achieving stuck
at non-optimal policies [33], [34]. Hence, the action selection
probability of the learning agent is expressed as
\begin{equation}
\begin{split}
{\rm{Pr}}(a = \tilde a) = \left\{ \begin{array}{l}
{\rm{1}} - \varepsilon ,\;\;\;\;\tilde a = \arg \mathop {\max }\limits_{a \in \mathcal{A}} Q(s,a),\\
\frac{\varepsilon }{{|\mathcal{A}| - 1}},\tilde a \ne \arg \mathop {\max
}\limits_{a \in \mathcal{A}} Q(s,a).
\end{array} \right.
\end{split}
\end{equation}

As the WoLF-PHC algorithm is capable of not only keeping the
Q-function but also quickly learning the decision-making policy under uncertain characteristics [35], so this study adopts it to derive the optical power allocation and reflecting beamforming
strategy with the unknown jamming model.
In the IRS-assisted communication system, the jammer attempts to
disturb the communication performance of legitimate UEs by
exploiting the action of the learning agent, the WoLF-PHC
algorithm can provide uncertainty in the action selection and
fools the jammer's attacks in the presence of the unknown   jamming model.

In WoLF-PHC, the mixed policy $\pi (s,a)$ is updated by increasing
the probability that it selects the most valuable action with the
highest Q-function value by a learning rate  $\xi \in (0,1]$, and
reducing other probabilities by $ - \xi /(|\mathcal{A}| - 1)$, i.e.,
\begin{equation}
\begin{split}
\begin{array}{l}
\pi (s,a) \leftarrow \pi (s,a)\\
 + \left\{ \begin{array}{l}
\xi ,\;\;\;\;\;\;\;\;{\rm{if}}\;a = \arg {\max _{a'}}Q(s,a'),\\
 - {\textstyle{\xi  \over {|\mathcal{A}| - 1}}},\;{\rm{otherwise}}{\rm{.}}
\end{array} \right.
\end{array}
\end{split}
\end{equation}

In (14), the WoLF-PHC algorithm has two variable learning rates,
i.e.,  ${\xi _{{\rm{win}}}}$ and  ${\xi _{{\rm{loss}}}}$, with
${\xi _{{\rm{loss}}}} > {\xi _{{\rm{win}}}}$ [35]. These two learning
rates are employed to improve the learning policy depending on
whether the learning agent  wins or loses at the current learning step. The
expected value is generally used to take this determination, in
other words, whether the expected value of the current
policy  $\pi (s,a)$ is larger than the expected value of the
average policy
$\mathord{\buildrel{\lower3pt\hbox{$\scriptscriptstyle\frown$}}
\over \pi } (s,a)$. If the current expected value is larger (i.e.
the agent is ``winning"), then the smaller learning rate ${\xi
_{{\rm{win}}}}$ is chosen; otherwise, ${\xi _{{\rm{loss}}}}$ is
selected.  The corresponding updating rule is given by
\begin{equation}
\begin{split}
\xi  = \left\{ \begin{array}{l} {\xi
_{{\rm{win}}}},\;{\rm{if}}\;\sum\nolimits_a {\pi (s,a)Q(s,a)
> \sum\nolimits_a
{\mathord{\buildrel{\lower3pt\hbox{$\scriptscriptstyle\frown$}}
\over \pi } (s,a)Q(s,a)} }, \\
{\xi _{{\rm{loss}}}},\;{\rm{otherwise}}.
\end{array} \right.
\end{split}
\end{equation}

  When the selected action $a$ is executed, the average policy  $\mathord{\buildrel{\lower3pt\hbox{$\scriptscriptstyle\frown$}}
\over \pi } (s,a)$ of all actions is updated under the given state
$s$ as
\begin{equation}
\begin{split}
\mathord{\buildrel{\lower3pt\hbox{$\scriptscriptstyle\frown$}}
\over \pi } (s,a) \leftarrow
\mathord{\buildrel{\lower3pt\hbox{$\scriptscriptstyle\frown$}}
\over \pi } (s,a) + \frac{{\pi (s,a) -
\mathord{\buildrel{\lower3pt\hbox{$\scriptscriptstyle\frown$}}
\over \pi } (s,a)}}{{C(s)}}
\end{split}
\end{equation}
where $C(s)$ denotes the number of the state $s$ from the
initial state to the current state.

In practical communication systems, it is difficult to know the
accurate jamming model and behaviors of the smart jammer, so the
possible estimated state space related to the jamming behaviors
increases. Moreover, the uncertain jamming behaviors and mobility
of the jammer result in an uncertain dynamic during the learning
process. Hence, this paper combines FSA into the WoLF-PHC
algorithm to represent the system states with a fixed number of
FSA states and thus reduce the number of system states that
the WoLF-PHC algorithm must deal with. FSA can transform
continuous states as discrete [36], hence enabling the use of
 the WoLF-PHC algorithm in continuous state spaces. Thus,
FSA is combined with the WoLF-PHC algorithm to formulate a fuzzy
approximation architecture, where the fuzzy state-action value
function under the state-action pair ($s$, $a$) is given by
\begin{equation}
\begin{split}
FQ(s,a) = \sum\limits_{l = 1}^L {{Q_l}(s,a)} {\psi _l}(s,a)
\end{split}
\end{equation}
where $L$ represents the number of  fuzzy states,  ${FQ_l}(s,a)$ is
the value function of the $l$-th fuzzy state and  ${\psi _l}(s,a)$
stands for the degree of relationship between the state
$s$ and the $l$-th fuzzy state with the given action $a$.

The mixed policy  $\pi (s,a)$ controls the probabilities which is
utilized to choose an action with a given policy. With FSA, the
policy decision is determined by both the expected reward value
${Q_l}(s,a)$ and the policy ${\pi _l}(s,a)$, i.e.,
\begin{equation}
\begin{split}
{\pi _{{\rm{FSA}}}}(s,a) = \sum\limits_{l = 1}^L {{\pi
_l}(s,a){Q_l}(s,a)} {\psi _l}(s,a).
\end{split}
\end{equation}

And the elements of $\left\{ {{\pi _l}(s,a)} \right\}_{l = 1}^L$
are initialized by
\begin{equation}
\begin{split}
\sum\limits_{a = 1}^{|\mathcal{A}|} {\sum\limits_{l = 1}^L {{\pi
_l}(s,a)} {\psi _l}(s,a)}  = 1.
\end{split}
\end{equation}

The policy  ${\pi _l}(s,a)$ is updated as follow
\begin{equation}
\begin{split}
{\pi _l}(s,a) \leftarrow {\pi _l}(s,a) + \left\{ \begin{array}{l}
\frac{{\xi {\psi _l}(s,a)}}{L},\;\;\;\;{\rm{if}}\;a = \arg {\max _{a'}}FQ(s,a')\\
 - \frac{{\xi {\psi _l}(s,a)}}{{L(|\mathcal{A}| - 1)}},\;{\rm{otherwise}}{\rm{.}}
\end{array} \right.,\;\forall l.
\end{split}
\end{equation}

As WoLF-PHC is combined with FSA, the probability  $\xi $ is used
to update the entire fuzzy summation
\begin{equation}
\begin{split}
\sum\limits_{a = 1}^{|\mathcal{A}|} {\sum\limits_{l = 1}^L {{\pi
_l}(s,a)} {\psi _l}(s,a)}  \leftarrow \sum\limits_{a = 1}^{|A|}
{\sum\limits_{l = 1}^L {{\pi _l}(s,a)} {\psi _l}(s,a)}  + \left\{
\begin{array}{l}
\xi ,\;\;\;\;{\rm{if}}\;a = \arg {\max _{a'}}FQ(s,a'),\\
 - \frac{\xi }{{|\mathcal{A}| - 1}},\;{\rm{otherwise}}{\rm{.}}
\end{array} \right.
\end{split}
\end{equation}

It is worth pointing out  that with a given action $a$, the
probability $\xi $ needs to be carefully set in (20) and (21),  to avoid a disproportionate growth in ${\pi _l}(s,a)$ [36].

Accordingly, the update on the fuzzy state-action value function
is given by
\begin{equation}
\begin{split}
\begin{array}{l}
FQ(s,a) \leftarrow (1 - \alpha )FQ(s,a)\\
\;\;\;\;\;\;\;\;\;\;\;\;\; + \alpha \left( {r + \gamma \mathop
{\max }\limits_{a \in \mathcal{A}} F{Q^*}(s',a)} \right).
\end{array}
\end{split}
\end{equation}

The fuzzy WoLF-PHC-based joint power allocation and reflecting
beamforming approach for the IRS-assisted communication system
against smart jamming is summarized  in \textbf{Algorithm 1}. In the
system, at each episode training step, the learning agent observes
its system state  ${s^t}$ (i.e., the estimated jamming power, SINR
values, and channel coefficients) by interacting with the
environment. At each learning time slot $t$, the joint action
${a^t}$ (i.e., power allocation and reflecting beamforming) is
selected by using the probability distribution  $\pi
({s^t},{a^t})$. The $\varepsilon $-greedy policy method is
employed to balance the exploration and the exploitation,
for example, the action with the maximum Q-function value is chosen
with probability $1 - \varepsilon $ according to the known
knowledge, while a random action is chosen with probability
$\varepsilon $  based on the unknown knowledge. After executing
the selected action ${a^t}$, the environment will feedback a
 reward $r({s^t},{a^t})$ and a new system state ${s^{t +
1}}$ to the learning agent. Then, the WoLF-PHC algorithm updates both the
current policy $\pi ({s^t},{a^t})$ and the average policy
$\mathord{\buildrel{\lower3pt\hbox{$\scriptscriptstyle\frown$}}
\over \pi } ({s^t},{a^t})$, and uses them to select the variable
learning rate  $\xi $ to improve the learning rate. According to
the updated policy  $\pi ({s^t},{a^t})$, FSA is adopted to
calculate the fuzzy Q-function $FQ({s^t},{a^t})$ and the
fuzzy-state policy  ${\pi _{{\rm{FSA}}}}(s,a)$, as well as updating
$FQ({s^t},{a^t})$ in the next time slot until it converges to the
final point. Finally, the learning model is trained successfully,
and it can be loaded to search the joint power allocation ${\{
{P_k}\} _{k \in \mathcal{K}}}$  and reflecting beamforming matrix
${\bf{\Phi }}$ strategies according to the selected action.

\linespread{1.45}{
\begin{algorithm}[t]
\begin{normalsize}

\caption{\normalsize Fuzzy WoLF-PHC-Based Joint Power Allocation
and Reflecting Beamforming}

1:$~$\textbf{Input:} Fuzzy WoLF-PHC learning structure and IRS-assisted system with a jammer.\\
2:$~$\textbf{Initialize:}  $Q(s,a) = 0$, $\pi (s,a) = {1
\mathord{\left/
 {\vphantom {1 {|\mathcal{A}|}}} \right.
 \kern-\nulldelimiterspace} {|\mathcal{A}|}}$, $\mathord{\buildrel{\lower3pt\hbox{$\scriptscriptstyle\frown$}}
\over \pi } (s,a) = \pi (s,a)$, $\xi $, ${\xi _{{\rm{loss}}}} > {\xi _{{\rm{win}}}}$, $C(s)$, $\gamma $, $\alpha $, and set fuzzy rules.\\
3: $~$\textbf{for} each episode $j$  $=$ 1, 2, \dots,  ${N^{{\rm{epi}}}}$ \textbf{do}\\
4: $~~$ \textbf{for} each time step $t$ $=$ 0, 1, 2, \dots, $T$ \textbf{do}\\
5: $~~~$ Observe an initial system state  ${s^t}$;\\
6: $~~~$ Select an action $a^t$ based on the $\varepsilon $-greedy
policy via (13): \\
$~~~~~~~~~$ ${a^t} = \arg \mathop {\max }\limits_{{a^t} \in \mathcal{A}}
Q({s^t},{a^t})$,  with probability 1-$\varepsilon $;\\
$~~~~~~~~~$ ${a^t} = {\rm{random}}{\{ {a_i}\} _{{a_i} \in
\mathcal{A}}}$, with probability  $\varepsilon $;\\
7: $~~~$  Execute the exploration action ${a^t}$, receive a
reward $r({s^t},{a^t})$ and the next state ${s^{t + 1}}$;\\
8: $~~~$  Update the current policy $\pi ({s^t},{a^t})$ via
(14);\\
9: $~~~$ Select the variable learning rate $\xi $ via (15);\\
10: $~~$  Update the average policy
$\mathord{\buildrel{\lower3pt\hbox{$\scriptscriptstyle\frown$}}
\over \pi } ({s^t},{a^t})$ via (16);\\
11: $~~$ Set $C(s) = C(s) + 1$;\\
12: $~~$ Compute the fuzzy Q-function $FQ({s^t},{a^t})$ via (17);\\
13: $~~$ Update the fuzzy-state policy ${\pi _{{\rm{FSA}}}}(s,a)$ via (20);\\
14: $~~$  Update $FQ({s^t},{a^t})$ by via (22);\\
15: $~~$ \textbf{end for}\\
16: $~$ \textbf{end for}\\
17:  \textbf{Return:}  Fuzzy WoLF-PHC-based learning model;\\
18: \textbf{Output:} Load the learning model to achieve the joint power
allocation ${\{ {P_k}\} _{k \in \mathcal{K}}}$  and reflecting
beamforming matrix ${\bf{\Phi }}$ strategy.\\
\end{normalsize}
\label{alg_lirnn}
\end{algorithm}
}

\section{Simulation Results and Analysis}

\begin{figure}
\centering
\includegraphics[width=0.65\columnwidth]{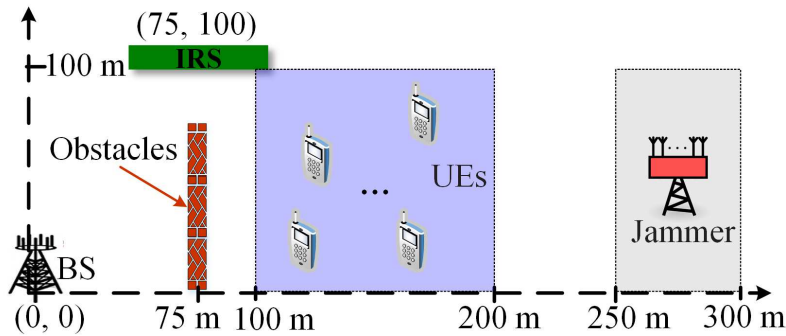}
\caption{{Simulation setup.} } \label{fig:Schematic}
\end{figure}

 This section evaluates the
performance of the IRS-assisted communication system against smart
jamming shown in Fig. 3, where a number of single-antenna UEs are
randomly located in the 100 m $\times$ 100 m right-hand side
rectangular area (light blue area). The locations of the BS and
the IRS are (0, 0) and (75, 100) in meter (m), respectively. There
are obstacles which block the direct communication links
from the BS to the UEs, so the obstacles cause a large-scale pathloss for the communication links.
In addition, a smart jammer is randomly located in the 50 m $\times$
100 m rectangle area (grey area), since the location behavior of
the smart jammer is not easy to be caught and it may move randomly
near the UEs.

As for the communication channel coefficients, the path loss in dB
is expressed as
\begin{equation}
\begin{split}
PL = \left( {P{L_0} - 10\beta \; {{\log }_{10}}(d/{d_0})} \right)
\end{split}
\end{equation}
where $P{L_0}$ denotes the path loss at the reference distance
${d_0}$,  $\beta $ is the path loss exponent, and $d$ is the
distance from the transmitter to the receiver, respectively. Here,
we use ${\beta _{{\rm{bu}}}}$,  ${\beta _{{\rm{br}}}}$, ${\beta
_{{\rm{ru}}}}$, and ${\beta _{{\rm{ju}}}}$ to denote the path loss
exponents of the channel links between the BS and the UEs, between the BS
and the IRS, between the IRS and the UEs, and between the jammer and the
UEs, respectively. According to [23]-[26], we set $P{L_0} =
30\;{\rm{dB}}$ and ${d_0} = 1\;{\rm{m}}$. Since there are
extensive obstacles as well as scatterers in the channel links
from the BS to the UEs, we set the path loss exponent ${\beta
_{{\rm{bu}}}} = 3.75$. As the IRS is carefully located in
favorable location to provide a low path loss, then we set  ${\beta _{{\rm{br}}}} =
{\beta _{{\rm{ru}}}} = 2.2$. The smart jammer is located near the UEs to
disturb the communication performance of UEs, and its corresponding
path loss exponent can be set to ${\beta _{{\rm{ju}}}} = 2.5$.

 We set  the background noise at all
UEs  equal to  ${\delta ^2} =  - 105$ dBm. The number of
antennas at the BS and the jammer are set to $N = {N_{\rm{J}}} =
8$.  The maximum transmit power ${P_{\max }}$  at the
BS varies from 15 dBm to 40 dBm, the SINR target value for  UEs varies from   $SIN{R^{\min }} = 10\;{\rm{dB}}$ to  $SIN{R^{\min }} =
25\;{\rm{dB}}$, and the number of IRS elements
$M$ varies from 20 to 100 for different simulation settings. In
addition, the jamming power of the smart jammer ranges from 15 dBm
to 40 dBm according to its jamming behavior, and the BS cannot
know the current jamming power levels, but it can
estimate the previous jamming power levels according to the
historical channel quality.

The learning rate is set to  $\alpha  = 0.5$, the discount factor
is set to $\gamma  = 0.9$ and the exploration rate is set to
$\varepsilon  = 0.1$. The cost parameters  ${\lambda _1}$ and
${\lambda _2}$ in (7) are set to ${\lambda _1} = 1$ and ${\lambda
_2} = 2$ to balance the utility and cost. We set ${\xi
_{{\rm{loss}}}} = 0.04$ and ${\xi _{{\rm{win}}}} = 0.01$ [35], [36]. The
following simulation results are averaged over 500 independent
realizations.

In addition, we compare the proposed fuzzy WoLF-PHC-based
joint power allocation and reflecting beamforming approach
(denoted as fuzzy WoLF-PHC learning) with the following
approaches:

\begin{itemize}
\item The system rate maximization approach which jointly optimizes the
BS's transmit power allocation and the IRS's reflect beamforming
by fixing other parameters as constants, and an iterative
algorithm is used to update power allocation and reflect beamforming under QoS constraints, which is similar to the suboptimal solution [29]
(denoted as Baseline 1 [29]).

\item The popular fast Q-Learning approach [19], which is adopted to optimize the transmit power allocation and reflecting beamforming in IRS-assisted communication systems (denoted as fast Q-Learning [19]).

\item  The optimal transmit power allocation at the BS without IRS
assistance (denoted as optimal PA without IRS).
\end{itemize}

\subsection{Convergence Comparisons of Different Approaches}

We first compare the convergence performance of all approaches when
${P_{\max }} = 30\;{\rm{dBm}}$,  $K=4$, $M = 60$, and
$SIN{R^{\min }} = 10\;{\rm{dB}}$. It is observed that the system
 rate and SINR protection level of all approaches (except the
optimal PA approach) increase with the number of iterations. Fig. 4
also indicates that the proposed fuzzy WoLF-PHC  learning approach
accelerates the convergence rate, enhances the system  rate
and increases the SINR protection level compared with both the
fast Q-Learning approach and the Baseline 1 approach. This is  because  that the
proposed leaning approach adopts WoLF-PHC and FSA techniques to
increase the learning rate and to enhance the learning
efficiency against smart jamming, yielding a faster learning rate
under the dynamic environment. Among all approaches, the fast Q-Learning
requires the largest number of convergence iterations to optimize
the Q-function estimator, where the slow convergence may fail to
protect SINR performance against smart jamming in real-time
systems. Moreover, the optimal PA approach
without IRS has the fastest convergence speed, but it obtains the  worst performance among all approaches, because it does not employ an
IRS for system performance improvement and jamming resistance.

\begin{figure}
\centering
\includegraphics[width=0.55\columnwidth]{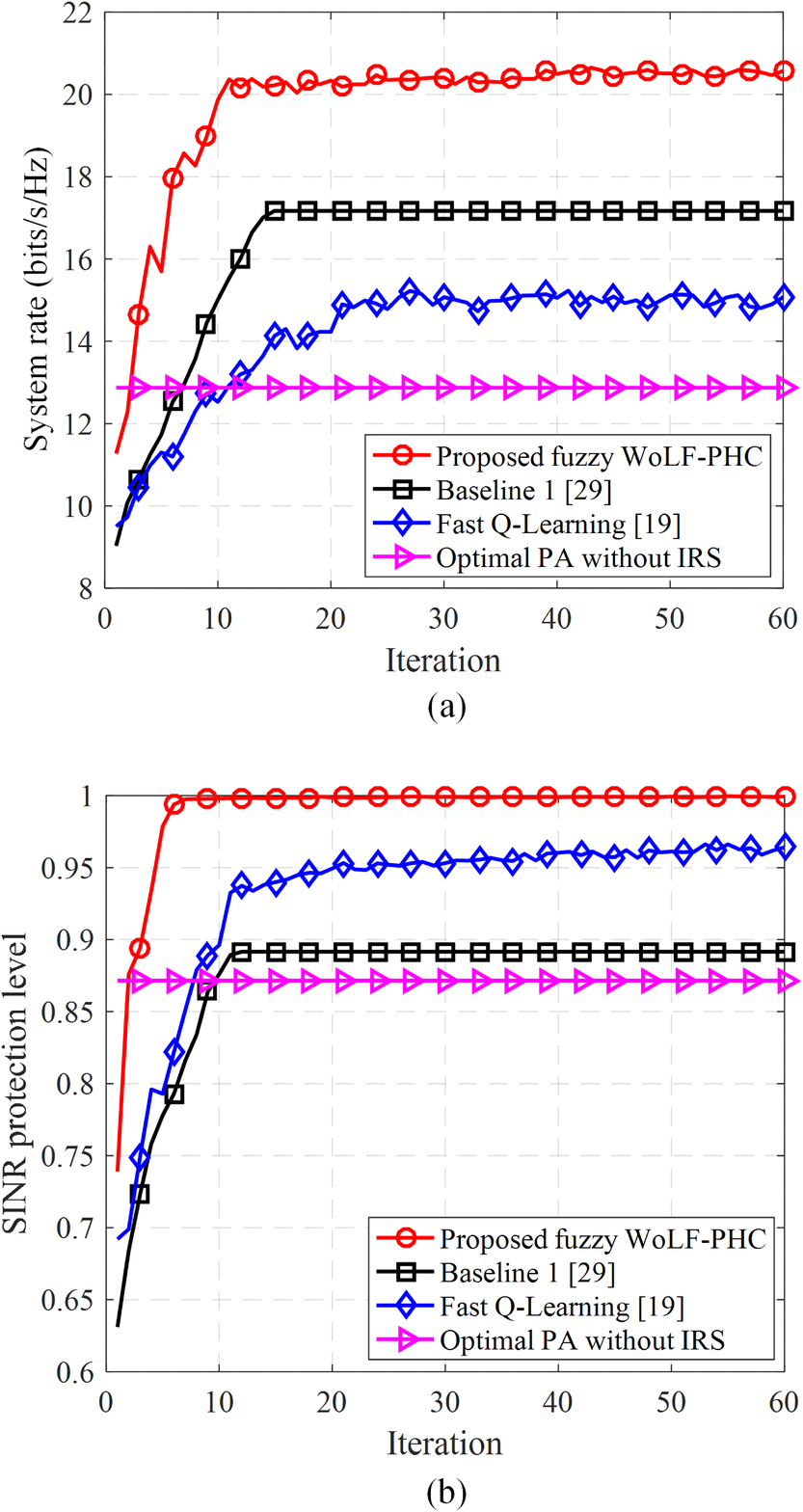}
\caption{{Convergence behaviors of the various approaches.} }
\label{fig:Schematic}
\end{figure}

As show in Fig. 4, the proposed learning approach is capable of
achieving a convergence within 11 iterations, while the fast
Q-Learning approach and the Baseline 1 approach require about 24
iterations and 15 iterations to achieve the convergence, respectively. In
addition, the proposed learning approach can improve the system
achievable rate by 21.29\% and increase the SINR protection level
by 13.36\% at a stable level, and it can also  save 36.36\% time to
converge to a stable value, compared with the Baseline 1 approach.
When all approaches finally achieve the convergence, the SINR
protection levels of all approaches are about 1, 0.89, 0.93 and
0.86, respectively. All these results demonstrate that our
proposed learning approach based on IRS assistance has the ability
to effectively improve the system performance and protect SINR
values against smart jamming.

\begin{figure}
\centering
\includegraphics[width=0.55\columnwidth]{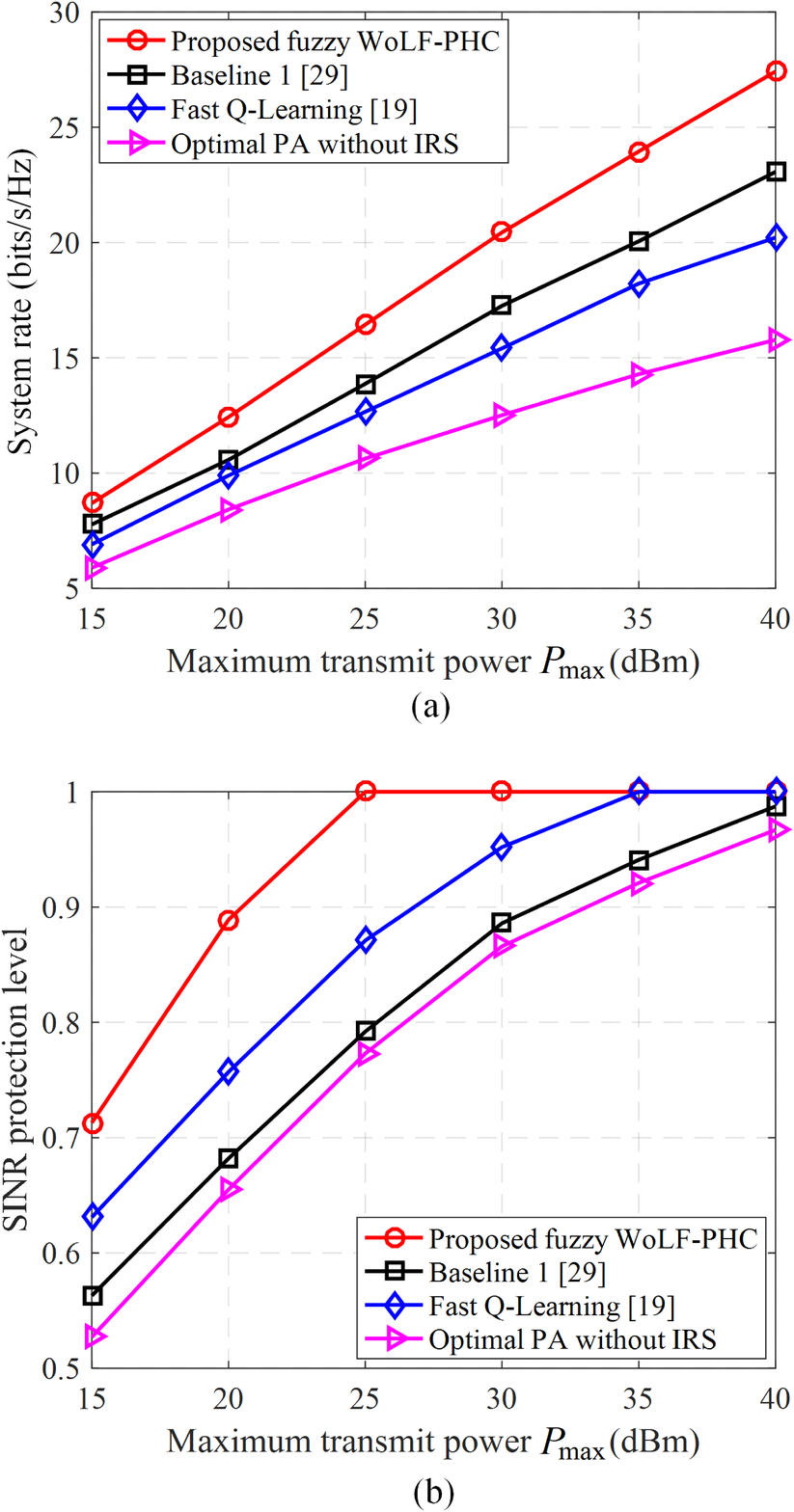}
\caption{{Performance comparisons versus the maximum transmit
power ${P_{\max }}$.} } \label{fig:Schematic}
\end{figure}

\subsection{Performance Comparisons versus Maximum Transmit Power}

The average system rate and SINR protection level versus the
maximum transmit power ${P_{\max }}$  for various approaches are
shown in Fig. 5, when  $K = 4$, $M = 60$, and  $SIN{R^{\min }} =
10\;{\rm{dB}}$, which demonstrates that the achieved system rate
and SINR protection level improve as  ${P_{\max }}$ increases. We can also
observe that both the proposed learning approach and the Baseline
1 approach have good system rate value under different values of  ${P_{\max }}$, and both of them greatly outperform other
approaches. However, the SINR protection level of the Baseline 1
approach is  obviously lower than that of the proposed learning
approach and fast Q-Learning approach, because it is a single time
slot optimization solution that  ignores  the long-term benefit,
and thus QoS requirement cannot be effectively guaranteed.

Additionally, the performance improvement achieved by using IRS versus no IRS increases  with  ${P_{\max }}$, which indicates the advantage of
deploying  IRS against smart jamming. In addition, the
performance of both the system rate and the SINR protection level
of the proposed fuzzy WoLF-PHC-based learning approach is higher
than that of the fast Q-Learning approach, which  is due to the fact
that WoLF-PHC and FSA  are adopted to effectively search the
optimal joint power allocation and reflecting beamforming strategy
against smart jamming in dynamic uncertain environments. From Fig.
5(b), it is interesting to observe that the optimal PA approach
without IRS has the comparable SINR protection level with the fast
Q-Learning approach based IRS-assisted system, but the optimal
approach is impractical  due to the assumption of the prefect CSI
and known jamming model in the system.

\subsection{Performance Comparisons versus Number of Reflecting Elements}

\begin{figure}
\centering
\includegraphics[width=0.55\columnwidth]{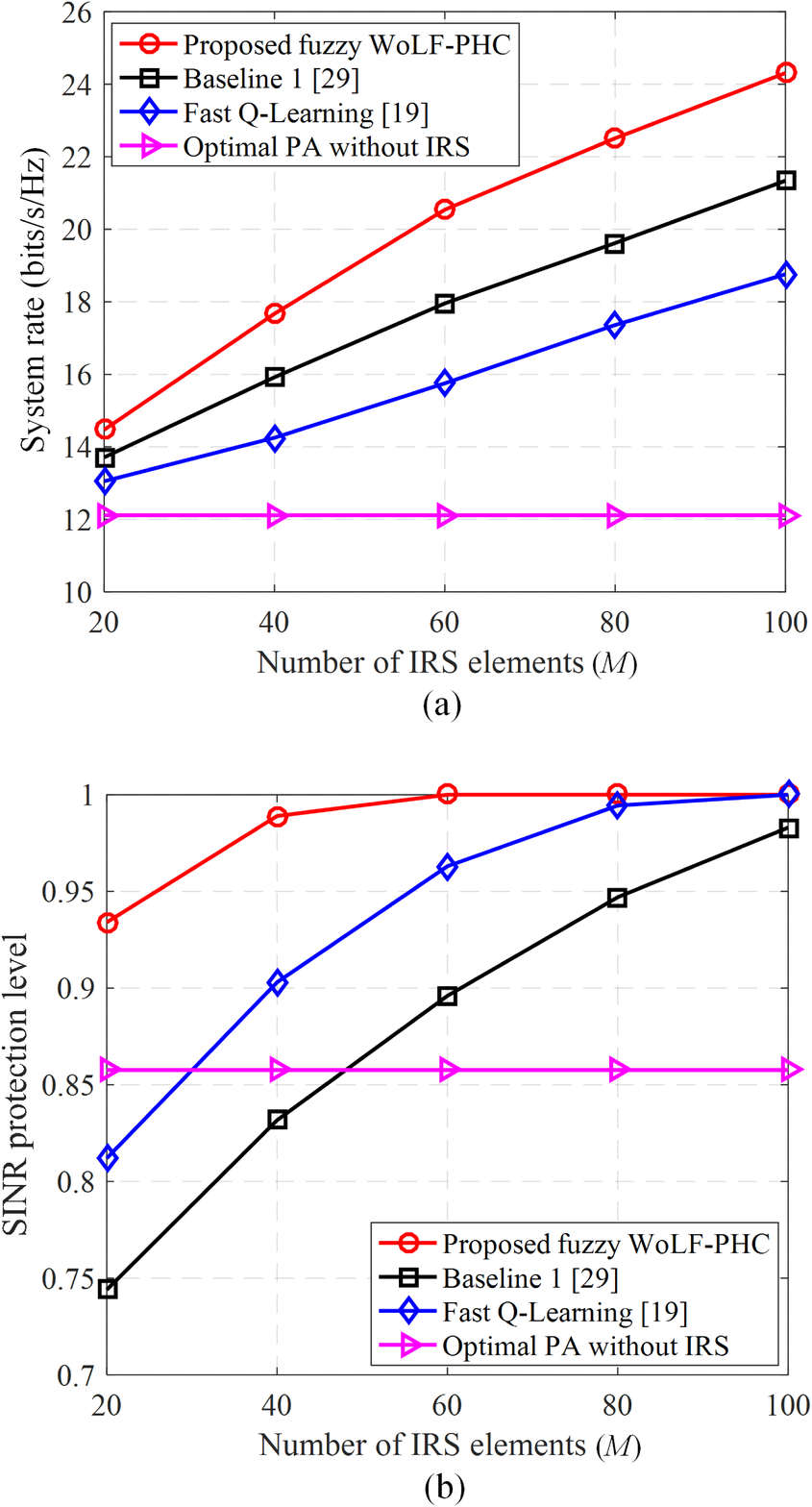}
\caption{{Performance comparisons versus the number of IRS
elements.}}  \label{fig:Schematic}
\end{figure}

Fig. 6 compares the performance of the four approaches for
different reflecting elements number $M$ when ${P_{\max }} =
30\;{\rm{dBm}}$,   $K = 4$ and $SIN{R^{\min }} = 10\;{\rm{dB}}$.
It can be seen that, except the optimal PA approach without IRS, the
performance of all IRS-based approaches increases with $M$,
and greatly outperforms the optimal PA approach without IRS. The reason is
 that the IRS has the ability to support higher degrees of freedom for
performance optimization, resulting in the great performance
gains obtained by employing the IRS against smart jamming over the
traditional system without IRS. Specifically, when  $M = 20$, the
system achievable rate gain of the proposed learning approach over
the optimal PA approach without IRS is only about 2.36 bits/s/Hz,
while this value is improved to 12.21 bits/s/Hz when  $M
= 100$. In addition, by deploying the IRS, the SINR protection
level is significantly improved compared with the optimal PA approach
without IRS assistance. Such performance improvement results
from the fact that higher power can be achieved at the IRS by
increasing $M$, and a higher reflecting beamforming gain is
achieved to design the IRS phase shifts to improve the received
desired signal as well as mitigate the jamming interference from
the smart jammer by increasing $M$.

In addition, from Fig. 6(a), we can also observe that the achievable
rate of the proposed learning approach outperforms both the fast
Q-Learning and Baseline 1 approaches, and  especially, the performance
gap  significantly increases with $M$. At the same time,  Fig. 6(b) shows that, as the reflecting elements
increases, the proposed learning approach is the first one can
achieve 100\% SINR protection level compared with other
approaches. This is because  deploying more reflecting
elements, the proposed fuzzy WoLF-PHC-learning based
joint power allocation and reflecting beamforming approach becomes
more flexible for optimal phase shift (reflecting beamforming)
design and hence achieves the higher performance gain. These
results also show that employing IRS into wireless
communications improves the anti-jamming communication performance
against smart jamming.

\subsection{Impact of SINR Target}

\begin{figure}
\centering
\includegraphics[width=0.55\columnwidth]{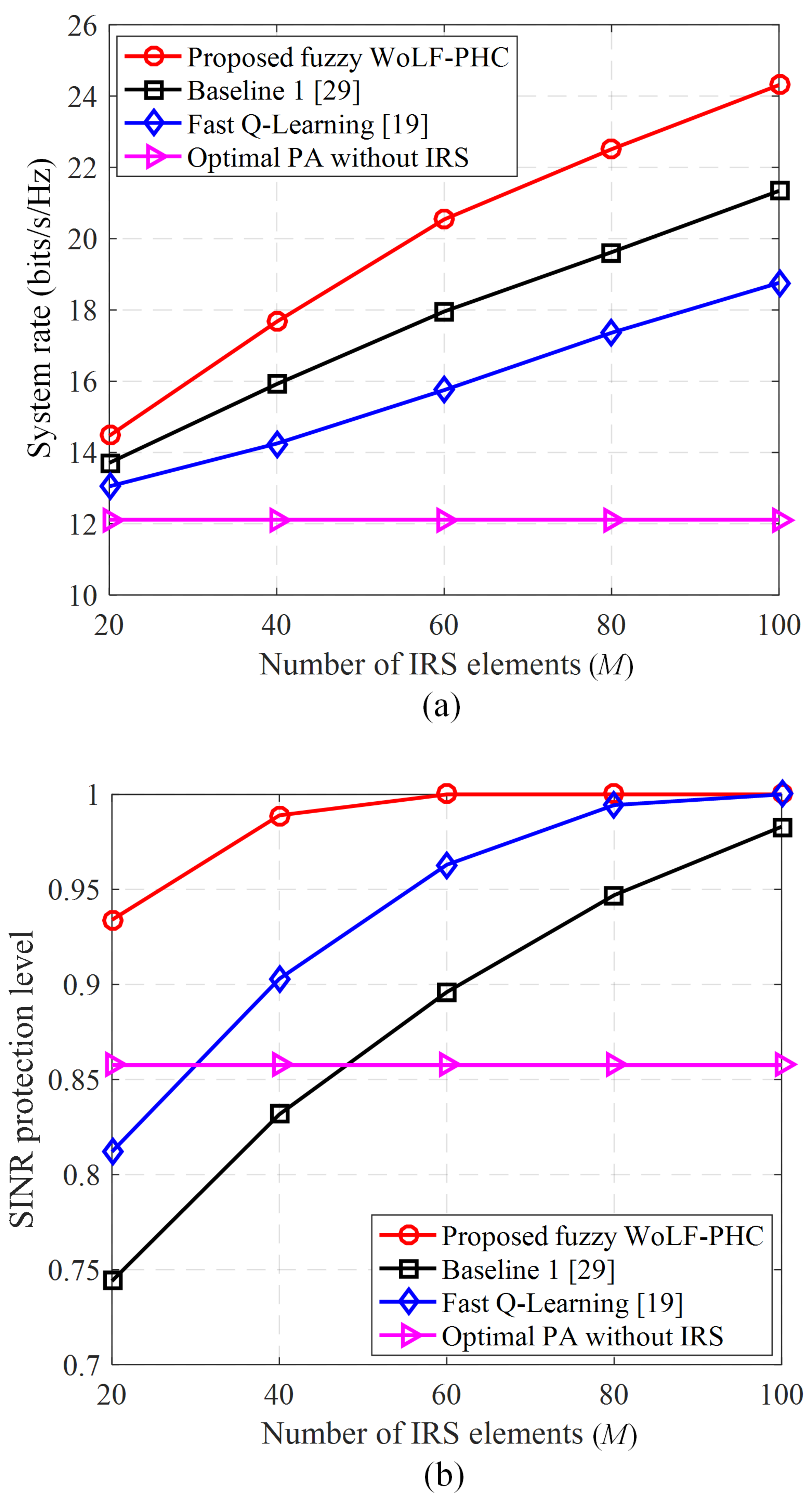}
\caption{{Performance comparisons versus UE SINR target.} }
\label{fig:Schematic}
\end{figure}

In Fig. 7, we investigate the effect of the UE SINR target
($SIN{R^{\min }}$) on the system performance for different
approaches when ${P_{\max }} = 30\;{\rm{dBm}}$, $K = 4$ , and $M =
60$. It can be seen from Fig. 7 that, for the low SINR target
regime, both the system rate and SINR protection level decline
slightly for increasing SINR target value, but the
performance drops significantly when $SIN{R^{\min }}$ is larger than a certain
threshold. The reason  is that that all approaches can still operate
an optimized power allocation and reflecting beamforming design to
maintain a favorable rate and guarantee the SINR requirement as
the  SINR target value is small, but the IUI and jamming
interference become the performance bottleneck when the UE SINR
target is high; especially, the jamming interference from the
smart jammer is out of control which mainly limits the performance
maintenance when the SINR constraint is stringent.

It is worth noting  that the three approaches based on
the IRS-assisted system achieve quite higher system rate and SINR
protection level performance than that of the optimal PA approach
without IRS, and the SINR
protection performance gap between them increases
significantly for increasing  SINR target value. This
further demonstrates that the deployment of IRS with reflecting
beamforming design can effectively enhance the desired signal
power and mitigate the jamming interference generated from the
smart jammer. From Fig. 7(b), among the three approaches in the
IRS-assisted system, our proposed fuzzy WoLF-PHC-based leaning
approach and the fast Q-Learning approach achieve a significant
improvement of the SINR protection level against smart jamming
compared with the Baseline 1 approach. The reason is that the two
learning approaches have the particular design of the SINR-aware
reward function shown in (7) for SINR protection against smart
jamming. Furthermore, our proposed learning approach achieves the
best performance, which indicates that the
joint power allocation and  reflect beamforming design
is crucial for the anti-jamming communication performance
against smart jamming, which needs the use  of both the
practical reflecting beamforming  and more sophisticated optimization.

 \section{Conclusions}

This paper has proposed to improve the anti-jamming performance
of wireless communication systems by employing an IRS. With the
assistance of IRS, the signal received at legitimate UEs is enhanced
while the jamming signal generated from the smart jammer can be
mitigated.  Specifically, we  have formulated  an
optimization problem by joint optimizing both the transmit power
allocation at the BS and the reflecting beamforming at the IRS. As
the non-convex optimization problem is not easy to solve and the
jamming model is unknown, a fuzzy WoLF-PHC learning approach has been proposed to achieve the optimal anti-jamming
strategy, where WoLF-PHC and FSA are capable of quickly achieving
the optimal policy without knowing the jamming model in uncertain
environments. Simulation results have confirmed  that that the IRS-assisted
system significantly improves the anti-jamming performance, and
also verified the effectiveness of our proposed learning
approach in terms of improving system rate as well as service protection
level, compared to other existing approaches.

\end{document}